\documentclass[aps,twocolumn,showpacs]{revtex4}
\usepackage{graphicx}   
\usepackage{latexsym}   

\newcommand{\bold}[1]{\mbox{\boldmath $#1$}}    
\newcommand{\del}{\partial}                     
\newcommand{\GeV}{{\rm GeV}}			
\newcommand{\MeV}{{\rm MeV}}			
\newcommand{\fm}{{\rm fm}}                      
\newcommand{\eps}{\varepsilon}			
\newcommand{\bfr}{\bold{r}}			
\newcommand{\grad}{\bold{\nabla}}		
\newcommand{\EoS}{equation of state}

\begin{document}
~\


\title{Spinodal amplification of density fluctuations\\
in fluid-dynamical simulations of relativistic nuclear collisions}

\author{Jan Steinheimer and J{\o}rgen Randrup}

\affiliation{
Nuclear Science Division, Lawrence Berkeley National Laboratory,
Berkeley, California 94720, USA}

\date{October 9, 2012}

\begin{abstract}
Extending a previously developed two-phase equation of state,
we simulate head-on relativistic lead-lead collisions
with fluid dynamics, augmented with a finite-range term,
 and study the effects of the phase structure
on the evolution of the baryon density.
For collision energies that bring the bulk of the system
into the mechanically unstable spinodal region of the phase diagram,
the density irregularities are being amplified significantly.
The resulting density clumping may be exploited as a signal
of the phase transition, 
possibly through an enhanced production of composite particles.

\end{abstract}


\maketitle

Strongly interacting matter is expected to posses a rich phase structure.
In particular, compressed baryonic matter 
may exhibit a first-order phase transition 
that persists up to a certain critical temperature \cite{Stephanov:1998dy} 
and experimental efforts are underway 
to search for evidence of this phase transition 
and the associated critical end point \cite{RHIC-BES,CBM-book,NICA}.

For this endeavor to be successful, it is important to identify
observable effects that may serve as signals of the phase structure.
This is a challenging task because the colliding system is 
relatively small, non-uniform, far from global equilibrium,
and rapidly evolving, 
features that obscure the connection between experimental observables and
the idealized uniform equilibrium matter described by the equation of state.
Therefore, to understand how the presence of a phase transition 
may manifest itself in the experimental observables,
it is necessary to carry out dynamical simulations of the collisions
with suitable transport models.

A first-order phase transition introduces spinodal instabilities
\cite{PhysRep,Randrup:2003mu,Sasaki:2007db,Randrup:2009gp}
and the associated non-equilibrium dynamics
may generate observable fluctuations in the chiral order parameter 
\cite{Paech:2003fe,NahrgangPRC84}
and the baryon density
\cite{RandrupAPH22,KochPRC72,Mishustin:1998eq,Bower:2001fq}.
In order to explore this latter prospect,
we simulate nuclear collisions with finite-density fluid dynamics,
using a previously developed two-phase equation of state.
Fluid dynamics has the important advantage that
the equation of state appears explicitly,
contrary to most microscopic transport models
where it is often unknown or is cumbersome to determine.
Because it is essential to incorporate finite-range effects when seeking to 
describe the spinodal phase decomposition 
\cite{PhysRep,Randrup:2003mu,Randrup:2009gp,Randrup:2010ax},
we introduce a gradient term into the expression for the local pressure
in a non-uniform medium.
This term ensures that two coexisting bulk phases
will develop a diffuse interface and we calculate 
the associated temperature-dependent tension.
The gradient term also causes the dispersion relation for the collective modes
in the unstable phase region to exhibit a maximum,
as is a characteristic feature of spinodal decomposition \cite{PhysRep}.
Thus we obtain a transport model 
that has an explicitly known two-phase equation of state 
and that treats the associated physical instabilities
in a numerically reliable manner.
This is the first time a transport model with these key characteristics
has been developed for high-energy nuclear collisions.

As a first application of this novel tool, 
we simulate head-on collisions of lead nuclei at various energies.
For a certain window of collision energies, several GeV per nucleon,
the bulk of the system will reside within the spinodal region 
of the phase diagram for a sufficient time to allow 
the associated instabilities to enhance the initial density irregularities,
a phenomenon that was exploited previously 
to obtain experimental evidence of the nuclear liquid-gas phase transition
\cite{BorderiePRL86,PhysRep}.
As a quantitative measure of the effect,
we extract the moments of the density distribution
and compare the degree of enhancement with what would result
in the absence of the phase transition.

In order to obtain a suitable equation of state, 
we employ the method developed in Ref.\ \cite{Randrup:2009gp}.
Thus we work (at first) in the canonical framework and, for a given $T$,
we obtain the free energy density $f_T(\rho)$ in the phase coexistence region 
by performing a suitable spline between two idealized systems
(either a gas of pions and interacting nucleons 
or a bag of gluons and quarks) held at that temperature.
In Ref.\ \cite{Randrup:2009gp} the focus was restricted to
 subcritical temperatures, $T<T_{\rm crit}$,
so for each $T$ the spline points were adjusted
so the resulting $f_T(\rho)$ would exhibit a concave anomaly,
{\em i.e.}\ there would be two densities, $\rho_1(T)$ and $\rho_2(T)$,
for which the tangent of $f_T(\rho)$ would be common.
This ensures phase coexistence because then the chemical potentials 
$\mu_T=\del_\rho f_T(\rho)$ match, $\mu_T(\rho_1)=\mu_T(\rho_2)$,
as do the pressures, $p_T(\rho_1)=p_T(\rho_2)$.
In the present work, we have extended the \EoS\ to $T>T_{\rm crit}$
by using splines that have no concavity,
as is characteristic of single-phase systems.
After having thus constructed $f_T(\rho)$
for a sufficient range of $T$ and $\rho$,
we may obtain the pressure, $p_T(\rho)=\rho\partial_\rho f_T(\rho)-f_T(\rho)$,
and the energy density,
$\eps_T(\rho)=f_T(\rho)-T\partial_T f_T(\rho)$, by suitable interpolation 
and then tabulate the \EoS, $p_0(\eps,\rho)$, 
on a convenient cartesian lattice.

\begin{figure}[t]	
\includegraphics[angle=0,width=3.4in]{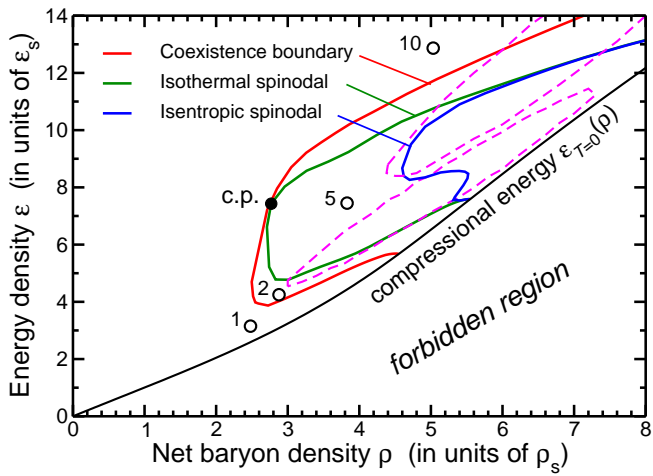}
\caption{[Color online]The $(\eps,\rho)$ phase diagram:
$\eps_{T=0}(\rho)$, the minimal $\eps$ for a given $\rho$ (black);
the phase-coexistence boundary (red) ending at the critical point (c.p.), 
as well as the isothermal spinodal boundary where $v_T\!=\!0$ (green)
and the isentropic spinodal boundary where $v_s\!=\!0$ (blue).
Points corresponding to counter-streaming Lorentz-contracted nuclei
are shown for various energies $E_{\rm lab}$, 
indicated in $A\,\GeV$ (circles).
The two contours (dashed) show where the points of maximum compression
are concentrated in ensembles of collisions 
simulated with fluid dynamics for $E_{\rm lab}$=2 or 5 $A\,\GeV$.
}\label{f:1}
\end{figure}

The resulting phase diagram in the $(\eps,\rho)$ phase plane
is illustrated in Fig.\ \ref{f:1}.
At a given net baryon density $\rho$,
the lower bound on the energy density is given by $\eps_{T=0}(\rho)$,
which represents the effect of compression.
The lower and upper boundaries of the phase coexistence region
are traced out by $\rho_1(T)$ and $\rho_2(T)$, respectively.
Uniform matter is thermodynamically unstable inside this region;
while it is mechanically metastable near the phase coexistence boundaries,
it looses mechanical stability when the speed of sound vanishes and 
density irregularities are amplified inside this spinodal boundary.

It is important to recognize that the model describes the instabilities 
not only in the unstable spinodal region,
but also those in the surrounding metastable region 
in which finite seeds are required for amplification to occur 
(yielding nucleation or bubble formation \cite{Mishustin:1998eq}).
Thus density irregularities may be amplified by the metastable region as well,
and any clumping generated inside the spinodal region may be further enhanced
as the system expands through the nucleation region.

When the dynamical evolution is governed by ideal fluid dynamics (see below)
the instability boundary is characterized by the vanishing
of the isentropic sound speed, $v_s=0$,
where $v_s^2\equiv(\rho/h)\left(\del p/\del\rho\right)_{s/\rho}$,
with $s$ being the entropy density and $h\!=\!p+\eps$ the enthalpy density.
For dissipative evolutions the finite heat conductivity causes the
instability region to widen as the boundary is then characterized by
the vanishing of the isothermal sound speed, $v_T=0$,
where $v_T^2\equiv(\rho/h)\left(\del p/\del\rho\right)_T\leq v_s^2$.
These instability boundaries are also indicated on Fig.\ \ref{f:1}.

In order to ascertain the dynamical effects 
of the first-order phase transition,
we construct a one-phase partner \EoS\ by replacing,
for any $T<T_{\rm crit}$, the actual $f_T(\rho)$
by the associated Maxwell construction,
$f_T^M(\rho)=f_T(\rho_i)+(\rho-\rho_i)\mu(\rho_i)$,
through the coexistence region $\rho_1(T)<\rho<\rho_2(T)$
where $f_T^M(\rho)<f_T(\rho)$.

For our present investigation, we describe the evolution of the
colliding system by ideal fluid dynamics,
because dissipative effects are not expected to play a
decisive role for the spinodal clumping \cite{Randrup:2010ax}.
The basic equation of motion expresses four-momentum conservation,
$\partial_\mu T^\mu=0$,
where $T^{\mu\nu}(x) = [p(x)+\eps(x)]u^\mu(x)u^\nu(x)-p(x)g^{\mu\nu}$
is the stress tensor.
It is supplemented by the continuity equation $\partial_\mu N^\mu=0$
for the baryon charge current density $N^\mu(x)=\rho(x)u^\mu(x)$,
where $u^\mu(x)$ is the four-velocity of the fluid.
These equations of motion are solved by means of
the code SHASTA \cite{Rischke:1995ir}
in which the propagation in the three spatial dimensions
is carried out consecutively.

A proper desciption of spinodal decomposition requires that
finite-range effects be incorporated
\cite{PhysRep,Randrup:2003mu}.
Therefore, following Refs.\ \cite{Randrup:2009gp,Randrup:2010ax},
we write the local pressure as 
\begin{equation}\label{a}
p(\bfr)	=p_0(\eps(\bfr),\rho(\bfr))
-a^2{\eps_s\over\rho_s^2}\rho(\bfr)\grad^2\rho(\bfr)\ ,
\end{equation}
where we recall that $p_0(\eps,\rho)$ is the \EoS,
the pressure in uniform matter characterized by $\eps$ and $\rho$.
Furthermore, $\rho_s=0.153/\fm^3$ is the nuclear saturation density
and $\eps_s\approx m_N\rho_s$ is the associated energy density.
The strength of the gradient term is then conveniently
governed by the length parameter $a$.

\begin{figure}[t]	
\includegraphics[angle=0,width=3.2in]{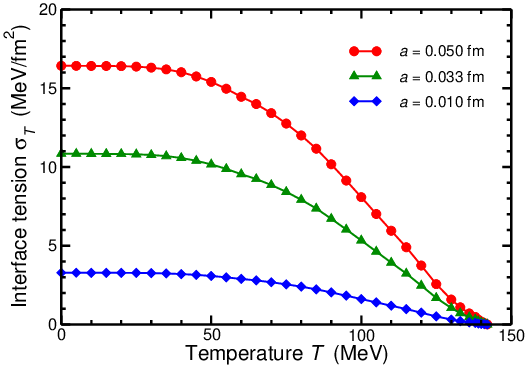}	
\includegraphics[angle=0,width=3.15in]{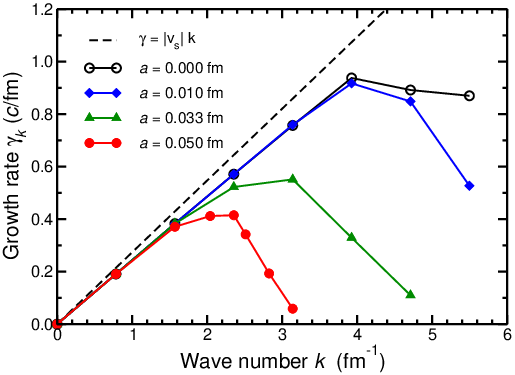}	
\caption{[Color online] Effect of the gradient term:
For various values of the parameter $a$ 
in Eq.\ (\ref{a}) is shown the interface tension $\sigma_T$ 
as a function of $T$ ({\em upper panel})
and the growth rate $\gamma_k$ as a function of the wave number $k$
of a disturbance, as obtained with ideal fluid dynamics ({\em lower panel}).
}\label{f:2}
\end{figure}

The gradient term will cause a diffuse interface to develop
when matter of two coexisting phases are brought into physical contact.
The associated interface tension
can readily be determined from the \EoS\ \cite{Randrup:2009gp},
\begin{equation}
\sigma_T\ 
=\ a\int_{c_1(T)}^{c_2(T)}\left[2\eps_s\Delta f_T(c)\right]^{1/2}dc\ ,
\end{equation}
where $c\equiv\rho/\rho_s$ denotes the degree of compression and
$\Delta f_T(c)\equiv f_T(c)-f_T^M(c)$ is the difference between the
actual free energy density at the specified compression
and that associated with the corresponding Maxwell construction.
The tension $\sigma_T$ decreases steadily as $T$ is raised
and finally vanishes at $T_{\rm crit}$; it is displayed
in Fig.\ \ref{f:2} as obtained for various values of the range $a$.

Furthermore, uniform matter inside the spinodal region (where $v_s^2<0$)
is mechanically unstable and density ripples of wave number $k$
will be amplified at a rate $\gamma_k(\rho,\eps)$
that depends on the strength of the gradient term,
$\gamma_k^2=|v_s|^2k^2-a^2(\eps_s/h)(\rho/\rho_s)^2k^4$.
Thus the gradient term introduces a penalty for the development
of short-range undulations and $\gamma_k$ exhibits a maximum
at the favored length scale;
the familiar dispersion relation for ideal fluid dynamics,
$\gamma_k=|v_s|k$, is recovered in the absence of a gradient term, $a=0$.

These spinodal growth rates have been extracted by following the
fluid-dynamical evolution of small harmonic perturbations in uniform matter.
Figure \ref{f:2} shows the resulting $\gamma_k$ curves
extracted at the phase point $(6\rho_s,10\eps_s)$
located in the central spinodal region (see Fig.\ \ref{f:1}).
Because the numerical algorithm inevitably produces some degree of dissipation
\cite{Rischke:1995ir},
the growth rates extracted for $a=0$
deviate from the exact ideal dispersion relation $\gamma_k=|v_s|k$.
Our studies suggest that while those results are reliable 
only up to $k\lesssim4/\fm$ the use of a finite range, $a>0$,
extends validty to significantly finer scales.

Having thus shown that our model produces 
both a meaningful interface tension
and reasonable spinodal growth rates,
we apply it to central collisions of lead nuclei
bombarded onto a stationary lead target
at various kinetic energies, $E_{\rm lab}$.
For each energy,
an ensemble of $200$ separate evolutions are generated,
each starting from a different initial state obtained by means of
the UrQMD model (in its cascade mode)
\cite{Bass:1998ca,Bleicher:1999xi,Petersen:2008dd}
in order to take account of the non-equilibrium dynamics 
in the very early stage of the collision. 
The switch from UrQMD to fluid dynamics occurs at a time $t_0$
when the two Lorentz-contracted nuclei have just passed through each other
and all initial collisions have occurred.
The needed fluid-dynamical quantities,
$\rho(\bfr)$, $\eps(\bfr)$, $\bold{u}(\bfr)$, 
are then extracted from the UrQMD state
by representing each hadron by a gaussian of $1$~fm width
and mapping this information 
onto a cartesian spatial lattice \cite{Steinheimer:2007iy}.
Thus the effects of stopping as well as local event-by-event fluctuations
are naturally included in the ensemble of the initial states.

Figure \ref{f:2} brings out the intimate relationship between the interface
tension $\sigma_T$ and the spinodal growth rates $\gamma_k$
and the range parameter $a$ merely presents a convenient 
means for linking the two properties.
For the present paper, we have adjusted $a$ so that the zero-temperature
interface tension is approximately 10~MeV/fm$^2$, 
the value that was also preferred in Ref.\ \cite{Mishustin:1998eq}.
This corresponds to the middle curves in Fig.\  \ref{f:2},
obtained for $a=0.033\,\fm$.

The key effect of the first-order phase transition
is the spinodal amplification of spatial irregularities 
\cite{Randrup:2003mu,Randrup:2009gp,Randrup:2010ax,PhysRep}.
In order to obtain a quantitative measure of the 
resulting degree of clumping in the system, we extract 
the moments of the (net) baryon density distribution $\rho(\bold{r})$,
\begin{equation}\label{moments}
\langle\rho^N\rangle\	\equiv\		
{1\over A^N}\int\rho(\bold{r})^{N}\rho(\bold{r})\,d^3\bold{r}\ ,
\end{equation}
where $A=\int\!\rho(\bold{r})d^3\bold{r}$ is the total (net) baryon number.
Thus the corresponding normalized moment,
$\langle\rho^N\rangle/\langle\rho\rangle^N$, is unity for $N=1$.
These quantities have observational relevance
because they are intimately related to 
the relative production yield of composite baryons.

\begin{figure}[t]	
\includegraphics[angle=0,width=3.2in]{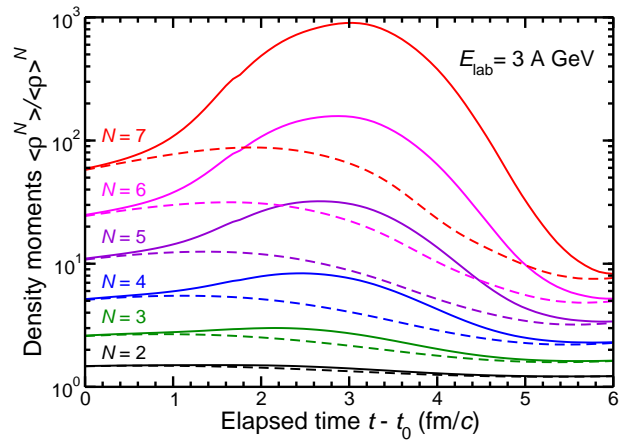}	
\caption{[Color online] Mean time evolution of normalized density moments
obtained for $E_{\rm lab}=3A\,\GeV$ with $a=0.033\,\fm$
(for which $\sigma_0\!\approx\!10\,\MeV/\fm^2$),
using either the two-phase equation of state (solid)
or its one-phase partner (dashed).
}\label{f:3}
\end{figure}		

The time evolution of the normalized density moments
is illustrated in Fig.\ \ref{f:3}.
Generally, for a given density distribution $\rho(\bfr)$,
the normalized moments increase with the order $N$, as one would expect.
At the initial time $t_0$ the pre-equilibrium fluctuations
are transcribed into the fluid-dynamical functions (see above)
and they may then be amplified subsequently 
by the instabilities associated with the first-order phase transition.
For comparison we show also the corresponding results
for the one-phase partner \EoS\ that has no instabilities.
There is a striking difference between the two sets of results:
Whereas the one-phase \EoS\ hardly produces any amplification at all,
the one having a phase transition leads to significant enhancements
that increase progressively with $N$
and amount to well over an order of magnitude for $N=7$.
This clearly demonstrates that the first-order phase transition
may have a qualitative influence on the dynamical evolution of the density.
Due to the variations in the initial conditions,
the phase regions explored differ from one collision to the other.
As a result, some of the evolutions experience amplifications
that are significantly larger than the ensemble average
(by up to a factor of five or so),
whereas more evolutions are affected considerably less.

Figure \ref{f:3} depicts what would happen
if the expansion were to continue while maintaining local equilibrium
so the generated density enhancements eventually subside.
However, as the hadronic gas grows more dilute, 
local equilibrium cannot be maintained.  Consequently,
if the decoupling occurs sufficiently soon after the clumps are formed,
the associated phase-space correlations may survive.
Studies of this issue are underway.

\begin{figure}[t]	
\includegraphics[angle=0,width=3.2in]{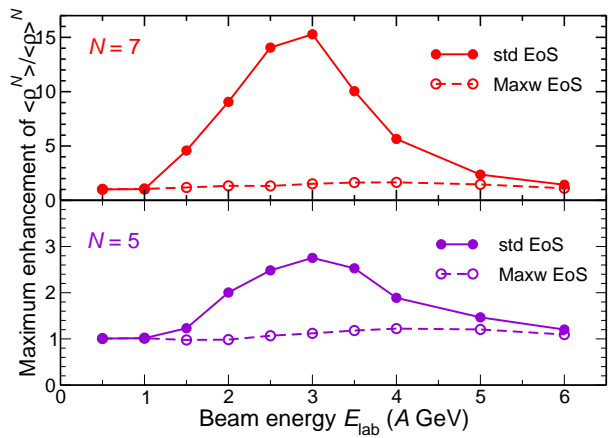}
\caption{[Color online] 
Mean maximum enhancement of the normalized density moments
for $N\!=\!7,5$ ({\em top, bottom})
as obtained for various energies
using either the two-phase equation of state (solid)
or its one-phase Maxwell partner (dashed).
}\label{f:4}
\end{figure}		

The degree of density clumping generated during a collision
depends on how long time the bulk of the matter is exposed to
the spinodal instabilities.
The optimal situation occurs for beam energies that produce maximum
bulk compressions lying well inside the unstable phase region
because the instabilities may then act for the longest time
\cite{Randrup:2009gp,Randrup:2010ax}.
At lower energies an ever smaller part of the system
reaches instability and the resulting enhancements are smaller.
Conversely, at higher energiers the maximum compression occurs
beyond the spinodal phase region and the system is exposed to the
instabilities only during a relatively brief period
during the subsequent expansion.
For still higher energies the spinodal region is being missed entirely.

Figure \ref{f:4} shows the (ensemble average) maximum enhancement achieved
as a function of the beam energy for the two equations of state.
The existence of an optimal collision energy is clearly brought out.
While the presently employed \EoS\ suggests that this optimal range is
$E_{\rm lab}\approx2-4\,A\,\GeV$, it should be recognized that others
may lead to different results.
Furthermore, the magnitude of the effect depends on the degree of fluctuation
in the initial conditions which in turn are governed by the pre-equilibrium
dynamics.
On the other hand, our studies suggest that the optimal energy
is rather insensitive to the range parameter $a$.

To summarize, by augmenting an existing fluid-dynamical code
with a gradient term, we have obtained a transport model 
that is suitable for simulating nuclear collisions 
in the presence of a first-order phase transition:
it describes both the tension between coexisting phases
and the dynamics of the unstable spinodal modes.
Applying this novel model to lead-lead collisions,
we have found that the associated instabilities may cause
significant amplification of initial density irregularities.
Because such clumping is expected to facilitate 
the formation of composite particles, such as deuterons and tritons,
this effect may be experimentally observable.

However, the magnitude of the generated clumping depends considerably
on the largely unknown specifics of the equation of state which determines 
whether the unstable region of the phase diagram is
entered during the collision,
for how long the system remains there,
and how rapidly irregularities are amplified.
It would therefore be interesting 
to expand the present kind of study to other equations of state
and also to ascertain the importance of dissipative effects 
\cite{Randrup:2010ax,Skokov:2009yu}.

Furthermore, standard fluid dynamics propagates the system deterministically
and thus ignores the effect of the ever present thermal noise.
Because the initial state, in the present study, 
is already endowed with significant fluctuations,
the thermal noise is likely to play only a minor role
(see Ref.\ \cite{PhysRep}), 
but it would be interesting to investigate this quantitatively.
(A study of the effect of noise in chiral fluid dynamics was recently made
\cite{Nahrgang}.)

Spinodal decomposition is an inherent characteristic
of first-order phase transitions and this 
non-equilibrium phenomenon may therefore provide especially compelling signals.
The present study demonstrates that the phase structure 
does affect the character of the density evolution
and we hope that these results will stimulate efforts
to develop analysis techniques for extracting the related observables
from experimental data.\\

\noindent{\bf Acknowledgments}\\
We wish to acknowledge stimulating discussion with Volker Koch.
This work was supported by the Office of Nuclear Physics
in the U.S.\ Department of Energy's Office of Science
under Contract No.\ DE-AC02-05CH11231;
JS was supported in part by the Alexander von Humboldt Foundation
as a Feodor Lynen Fellow.

		

			\end{document}